\documentclass[12pt]{article}
\usepackage{graphicx}
\begin{document}
\centerline{\bf Discrete Simulation of the Dynamics of Opinions about 
Extremism}

\bigskip
Dietrich Stauffer* and Muhammad Sahimi

\bigskip
\noindent
 
Department of Chemical Engineering, University of Southern California,
Los Angeles, California 90089-1211, USA

\bigskip
\noindent
Visiting from Institute for Theoretical Physics, Cologne University, 
D-50923 K\"oln, Germany.

\bigskip
e-mail: moe@usc.edu, stauffer@thp.uni-koeln.de

\bigskip
Abstract: {\small We propose a discrete model for how opinion about a given
phenomenon, about which various groups of a population have different degrees
of enthusiasm, such as fanaticism and extreme social and political positions, 
including terrorism, may spread. The model, in a certain limit, is the discrete
analogue of a deterministic continuum model suggested by others. We carry out 
extensive computer simulation of the model by utilizing it on lattices with 
infinite- or short-range interactions, and on (directed) Barab\'asi-Albert 
scale-free networks. Several interesting features of the model are 
demonstrated, and comparison is made with the continuum model.}

\section{Introduction}
Given the current political climate around the world, and the rise of extreme
ideologies everywhere, from the Middle East to Africa and Western Europe,
models that can provide insight into how such idealogies may spread in a 
society are clearly of great interest. In particular, given that, (1) the 
phenomenon of globalization has made interaction between people of various
nations around the globe much easier than two decades ago, and (2) the fact that
although extreme ideologies are usually advocated by very small fringe groups, 
but yet they continue to survive, it is important to understand the role of 
these two factors on the opinion about such antisocial behavior as terrorism.
The goal of the present paper is to suggest a model to study this problem, and
understand its implications.

Some simple models for terrorism or extreme opinions appeared years ago
in the physics \cite{galam} and sociological \cite{deffuant} literatures.
The present work was motivated by an article on bioterrorism \cite{bio}, but 
the methods that we describe and utilize can also apply to opinion dynamics 
regarding, for example, the latest ``Star Wars'' movie, fashion, a political
candidate running for office, or other questions and opinions with varying 
degrees of enthusiasm. Thus, we do not even try to define ``terrorism'' here, 
as the model that we consider is generic.

The population in our model consists of four parts, $G,\;S,\;E,$ and $F$
corresponding, respectively, to the general, susceptible, excited, and fanatic 
parts of the population. For simplicity, hereafter we use the same letters to
denote the fractions of the total population belonging to each group. Members 
of the population can be convinced by acquaintances from the $S,\;E,$ and $F$ 
groups to move from the $G$ group to $S$; from there by the $E$ and $F$ groups 
to change to $E$, and from there by $F$ to change to $F$. Moreover, members of
each of the three groups $S,\;E,$ and $F$ can change their status and go
back directly to the $G$ group. The dynamics of a model based on such a
partitioning of a population has been treated in the continuum limit \cite{bio} 
by deterministic nonlinear differential equations, depending only on the total 
fractions $G,\;S,\;E,$ and $F$. The continuum model can provide mathematically 
sufficient conditions for terrorism, or any other opinion about a certain
subject, to die out at long times, implying that in the long-time limit 
everybody will belong to the $G$ group , while the fractions $S,\;E,$ and $F$
shrink to zero which, when it comes to terrorism, is a good omen for the world.

However, as is well-known in the statistical physics of complex systems, 
deterministic continuum models represent mean-field approximations to the
actual problem which, although they allow for development of mathematical 
proofs for the existence or nonexistence of certain phenomena and provide us 
with a first guide, they are also unreliable. Such models cannot take into 
account the effect of fluctuations on the phenomena. In addition, such models
cannot take into account the effect of the internet, fax machines, and 
satellite television which have made long-range interactions between people on 
very large scale possible. For example, in a phenomenon somewhat close to the 
present problem, deterministic differential equations predicted \cite{shnerb} 
extinction, whereas proper discrete simulations on a square lattice did not
predict \cite{shnerb} the same phenomenon.

Therefore, the goal of the present paper is to carry out extensive simulation of
a discrete model opinion dynamics which, in a certain continuum limit, becomes
similar to a deterministic model proposed by Ref.\cite{bio}. We utilize
Monte Carlo simulations of a population of individuals. Such simulations may
be called agent-based outside physics, but are used in physics since half a 
century.

The plan of the paper is as follows. We first describe the deterministic
continuum model which is based on a set of nonlinear differential equations. We
then describe the discrete model which is developed by putting individuals on 
a two-dimensional lattice, but the individuals can still be influenced by all 
other individuals. Then, we restrict the influence to nearest neighbours. 
Finally, we replace the regular 2D lattice by a scale-free network of 
Barab\'asi-Albert type \cite{ba}. The main point of the paper is not testing
whether the model can provide quantitative predictions; rather, we deal only 
with the methods and how to implement them realistically. In particular we 
follow \cite{bio} in assuming that if $S=E=F=0$ at some moment, then these 
three quantities stay zero forever. Thus, one simulation corresponds to the 
opinion dynamics following one external event and does not include new 
external events to cause $S,\;E,$ and $F$ to become non-zero again.  

\begin{figure}[hbt]
\begin{center}
\includegraphics[angle=-90,scale=0.5]{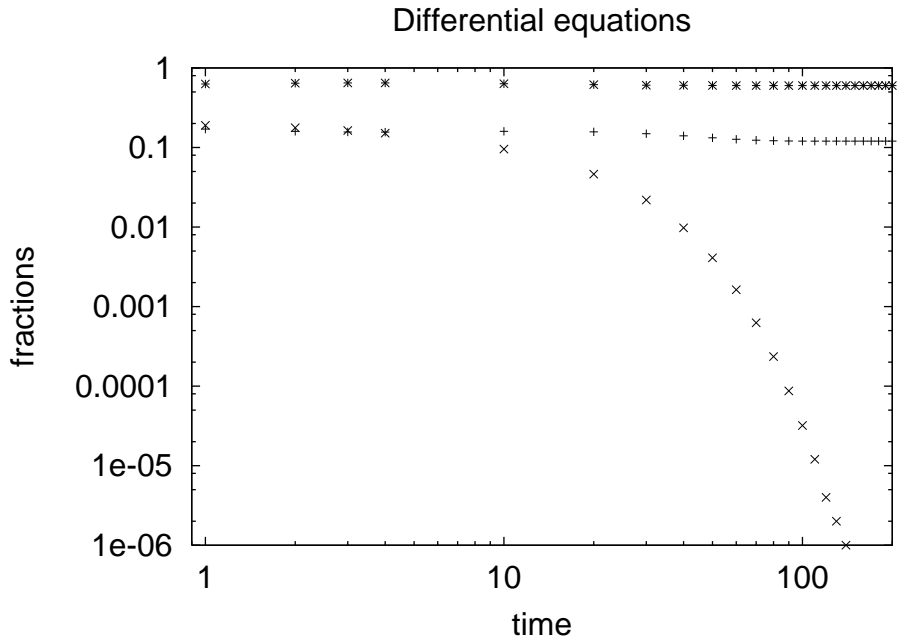}
\end{center}
\caption{
Differential equations:
Fractions of susceptible, excited and fanatic  opinions (from top to bottom)
at $\gamma_1 = 0.4$: Fanatics die out. For $\gamma_1 = 0.25$, fanatics remain;
for $\gamma_1 = 0.55$, excited and fanatic opinions die out, and for $\gamma_1
= 1.05$, also the susceptibles die out. 
}
\end{figure}

\section{The Deterministic Continuum Model}

The fractions $G,\;S,\;E,$ and $F$ in the population of agents having the 
corresponding opinions, with $C=S+E+F=1-G$, change with time $t$ as:
$$dS(t)/dt = \beta_1 CG - \beta_2 S(E+F)/C - \gamma_1S  \eqno (1a)$$
$$dE(t)/dt = \beta_2 S(E+F)/C - \beta_3 EF/C - \gamma_2E  \eqno (1b)$$
$$dF(t)/dt = \beta_3 EF/C - \gamma_3 E \quad .\eqno(1c)$$
Without loss of generality we set $\beta_1 = 1$ since, otherwise, it can be 
absorbed in the time scale. We also set $\gamma_2=\gamma_1$. Nevertheless, we 
still have not only the four parameters $\beta_2,\;\beta_3,\;\gamma_1,$ and
$\gamma_3$, but also the three initial concentrations $E(0),\;S(0),$ and 
$F(0)$, which are relevant due to the nonlinearity of the continuum model. We 
use, in general, one million people and $\beta_2=0.5,\;\beta_3=0.5,$ and 
$\gamma_3=0.20$, starting with $S=E=F=0.2$, which will be used throughout the 
paper. Then, we check for changes in the behaviour when we vary $\gamma_1$.

Figure 1 illustrates the behaviour: As predicted in \cite{bio}, for
$\gamma_1>\beta_1\;(=1$ here), only the general population remains;
for decreasing $\gamma_1$ first, $S$ also survives, then does $E$. Finally,
for $\gamma_1 = 0.25$ $F$ also survives, such that all the four groups,
$G,\;S,\;E,$ and $F$ remain present in the population.

\section{Averaged Lattice}
Now we put the agents onto a $1000\times 1000$ square lattice, half of whom
is selected randomly to be fixed as empty. The fractions $G,\;S,\;E,$ and $F$ 
now refer to the filled half of the lattice, i.e., they are fractions of the 
population and not of the lattice size and, thus, still add up to unity.
We try to follow closely the above set of equations, Eqs. (1), by the following
rules for each time step $t\to t+1$ (simultaneous updating of all agents):

\medskip
$G$ becomes $S$ with probability $\beta_1 C$;

$F$ becomes $G$ with probability $\gamma_3$;

$E$ becomes $G$ with probability $\gamma_2$ and $F$ with probability 
$\beta_3 F/C$, and

$S$ becomes $G$ with probability $\gamma_1$ and $E$ with probability 
$\beta_2(E+F)/C$.
\medskip

\noindent
These changes are simulated by first looking at the decay through $\gamma$
and then at the radicalisation through $\beta$. Therefore, it is possible that,
e.g., an $E$ first becomes $G$ and immediately thereafter changes opinion to 
$F$.

\begin{figure}[hbt]
\begin{center}
\includegraphics[angle=-90,scale=0.5]{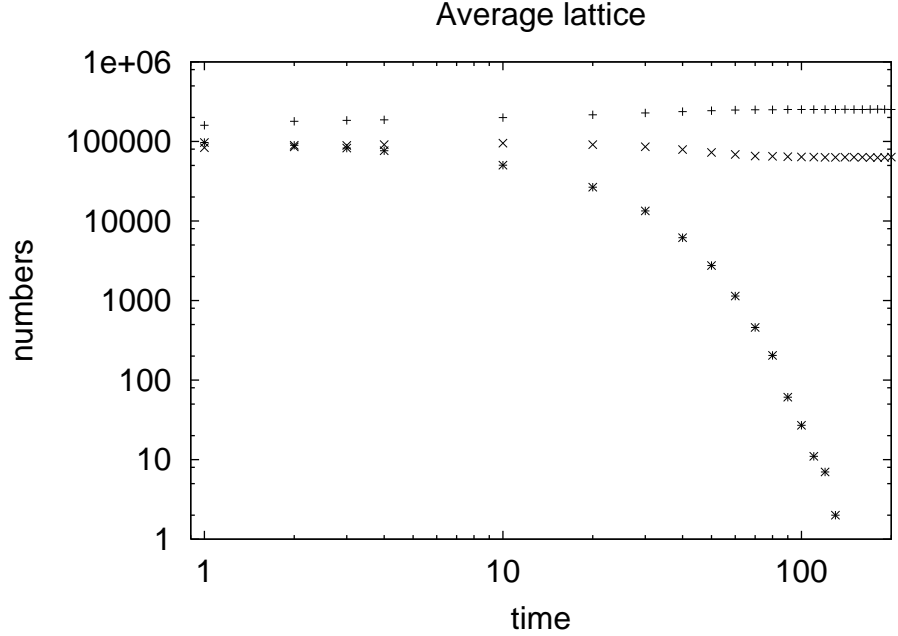}
\end{center}
\caption{
Averages on lattice:
Fractions of susceptible, excited and fanatic agents (from top to bottom)
at $\gamma_1 = 0.4$: Fanatics die out. For $\gamma_1 = 0.2$, fanatics remain;
for $\gamma_1=0.6$, excited and fanatic agents die out, but even for $\gamma_1
= 0.99$, some susceptibles remain.}
\end{figure}

Again we set $\beta_1=1,\;\gamma_2=\gamma_1$. 
Figure 2 looks similar to Fig. 1 which is not surprising since each agent is
affected by all other agents, which is the limit that in statistical physics 
is described by a mean-field approximation. However, we always find for 
probability $\gamma_1<1$ some susceptible people, which is in contrast to 
their extinction by continuum model. Only in the unrealistic limit $\gamma_1= 
1$ do they die out.

The reason for this persistence of susceptibles can be understood as follows:
For large enough $\gamma_3$, opinions $E$ and $F$ die out soon. Then,
we have the differential equation for $S=1-G$ as a simplification of Eq. (1a):
$$dS(t)/dt = \beta_1 (1-S)S - \gamma_1S \simeq (\beta_1-\gamma_1)S \eqno (2)$$
for small $S$, giving an exponential decay towards zero for $\gamma_1>\beta_1$.
For the Monte Carlo approach, $G$ becomes $S$ with probability $\beta_1 S$
and $S$ becomes $G$ with probability $\gamma_1$. Equilibrium thus requires,
for small $S$ and thus $G$ near unity, that:
$$ \gamma_1 S = G \beta_1 S = \beta_1(1-S)S \quad {\rm or} \quad 1-S = 
\gamma_1/\beta_1 \eqno (3)$$
which gives $S=1-\gamma_1$ for our choice $\beta_1 = 1$. Only for 
$\gamma_1 > \beta_1$ would $S$ become zero, which is not possible
if $\beta_1=1$ since $\gamma_1$ is a probability for the Monte Carlo approach
and no longer a rate which could also be larger than one. Putting back 
$\beta_1$ as a free parameter set equal to 0.5, everybody soon returns to
the general population, $S=E=F=0$, for $\gamma_1 = 0.6$. 

\begin{figure}[hbt]
\begin{center}
\includegraphics[angle=-90,scale=0.5]{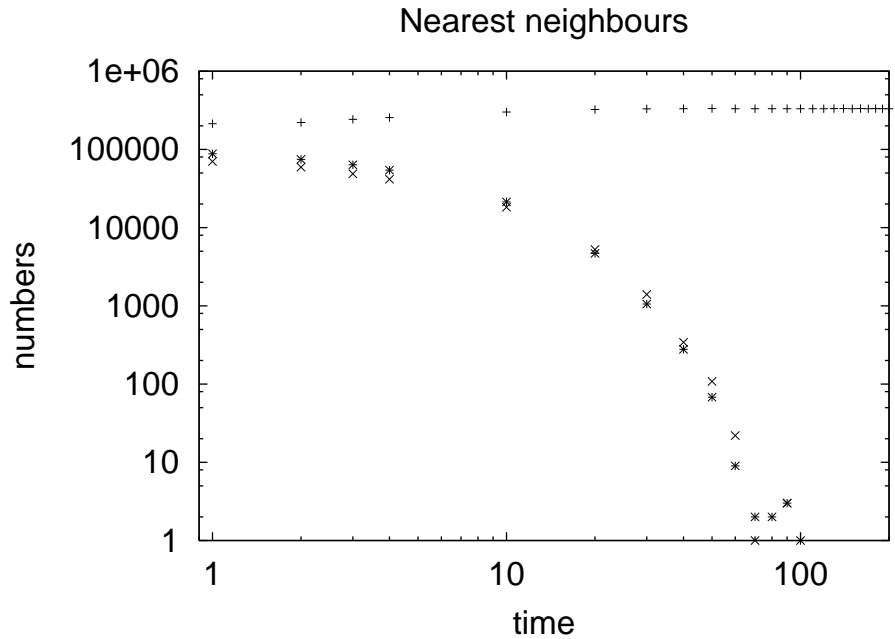}
\end{center}
\caption{Neighbour interactions on lattice:
Fractions of susceptible, excited and fanatic agents (from top to bottom)
at $\gamma_1 = 0.4$: excited and fanatic agents die out together. Both groups 
also die out at $\gamma_1 = 0.9$, while at $\gamma_1 = 0.1$ the fanatics 
die out soon whereas excited agents decay very slowly.}
\end{figure}

\section{Nearest Neighbour Interactions}

Now, we simulate a proper lattice population where only nearest neighbours
influence each other. Thus at each time step every agent selects randomly
one of the four nearest neighbours as a discussion partner and then follows
these rules (again $C = E+S+F$):

\medskip 

\noindent
$G$ becomes $S$ with probability $\beta_1$, if neighbour is  $S,\;E$, or $F$;

\noindent
$F$ becomes $G$ with probability $\gamma_3$;

\noindent
$E$ becomes $G$ with probability $\gamma_2$, and

it becomes $F$ with probability $\beta_3/C$, if neighbour is $F$;

\noindent
$S$ becomes $G$ with probability $\gamma_1$, and

it becomes $E$ with probability $\beta_2/C$, if neighbour is $E$ or $F$.

\medskip 
Thus, no agent is convinced by an empty neighbour to change opinion.
Again we set $\beta_1 = 1, \; \gamma_2 = \gamma_1$. Since the behaviour of the 
population now depends on the single opinions and not only on their sum over 
all lattice sites, ordered sequential updating with helical boundary 
conditions was used. 

Figure 3 shows that, differently from Figs.1 and 2, both $E$ and $F$, and not
only $F$, decay rapidly to zero; the susceptibles remain. 

\section{Networks}
Human beings are not trees in an orchard, sitting on a square lattice and 
having $k=4$ neighbours each. The previous section used a half-empty 
square lattice such that the number of neighbours varied between $k=0$ and 
$k=4$. In reality, some people have many friends and some only few. 
Such social relationships are much better described by scale-free \cite{ba}
networks of Barab\'asi-Albert type, where the number of people having
$k$ neighbours decays as $1/k^3$ for not too small $k$. Moreover, real 
terrorism obeys a power law \cite{clauset}.

To grow such a network we start with four people all connected with each
other and with themselves. Then, one after the other more sites are added,
and each new site selects four people of the existing network as
neighbours from whom to take advice. This selection is not random but
proportional to the current number $k$ of neighbours of that person:
Powerful people attract more ``friends`` than powerless people.

\begin{figure}[hbt]
\begin{center}
\includegraphics[angle=-90,scale=0.5]{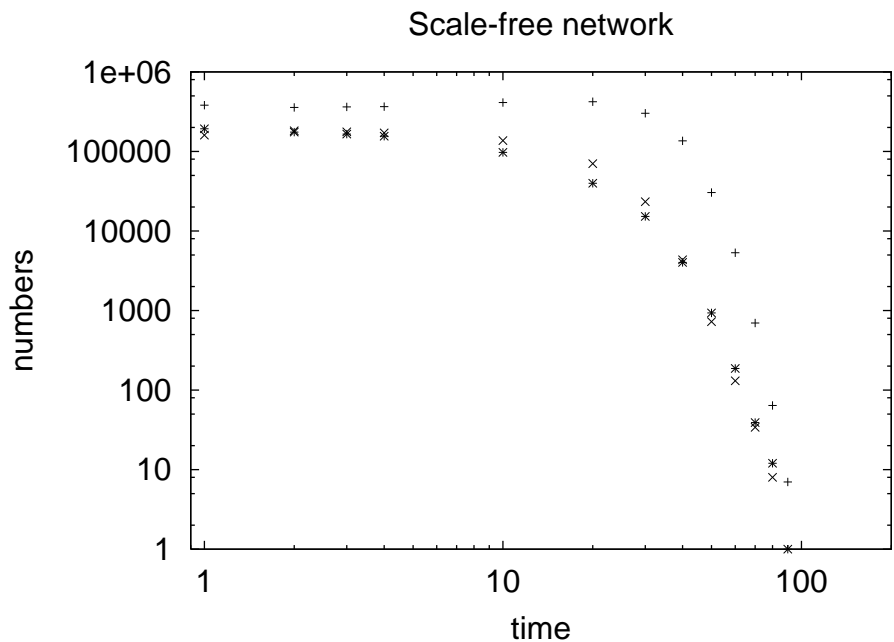}
\end{center}
\caption{Barab\'asi-Albert network: Now $S, E, F$ all die out for 
$\gamma_1=0.4$, 0.9 and (more slowly) 0.1.}
\end{figure}

\begin{figure}[hbt]
\begin{center}
\includegraphics[angle=-90,scale=0.5]{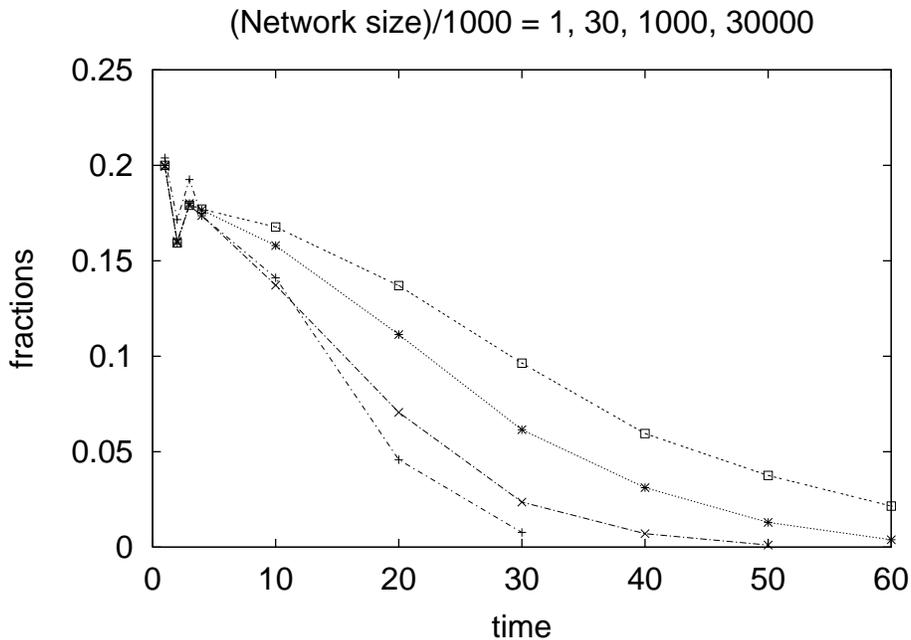}
\end{center}
\caption{Barab\'asi-Albert network: Decay of excited fraction $E$  becomes
slower for network size increasing from  1000 (leftmost curve) to 30,000, then 
1 million and finally (rightmost curve) 30 million. We averaged over 10 
samples.}
\end{figure}

\begin{figure}[hbt]
\begin{center}
\includegraphics[angle=-90,scale=0.5]{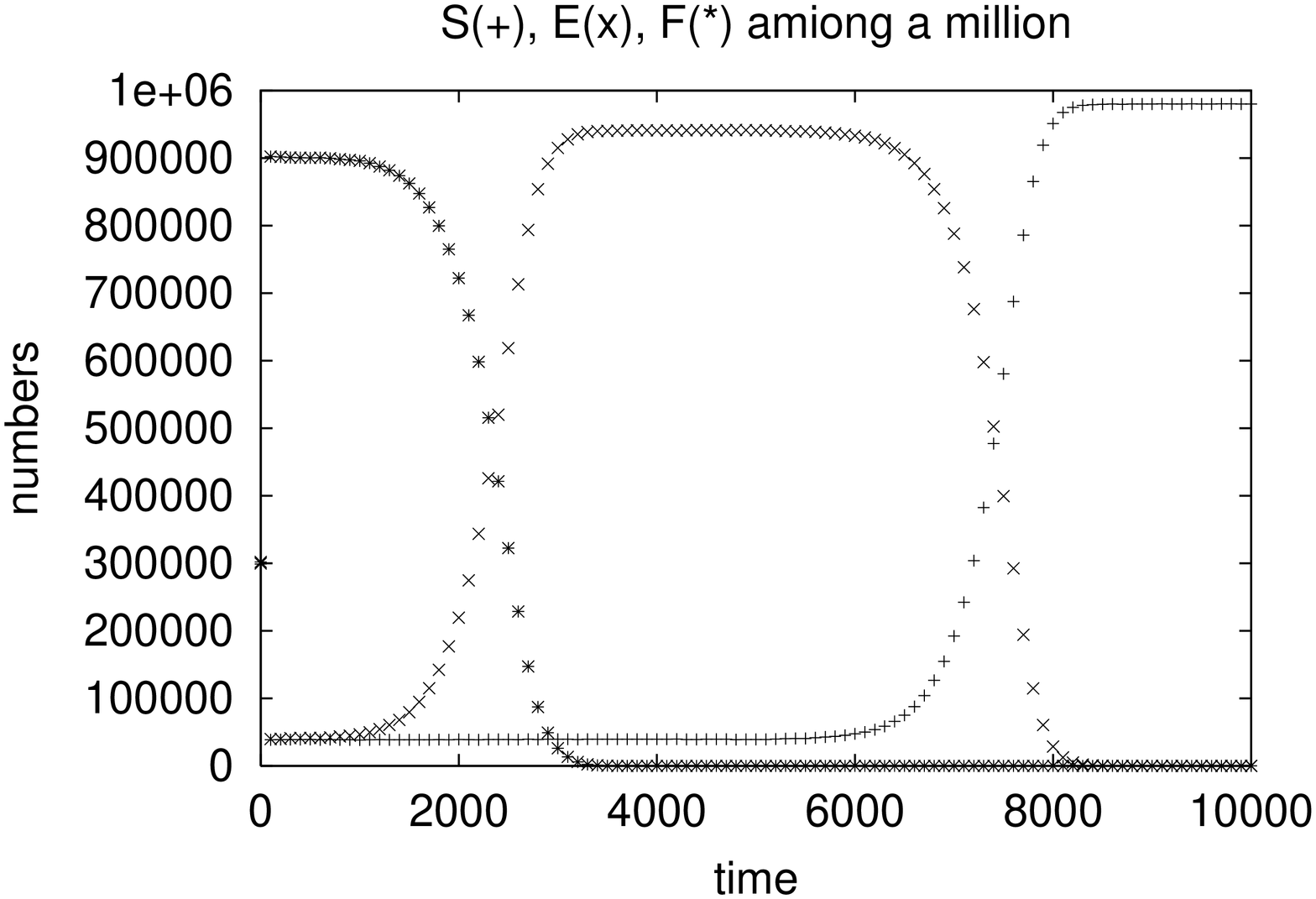}
\end{center}
\caption{Barab\'asi-Albert network:  Complicated dynamics at $\gamma_1=\gamma_2
=\gamma_3 = 0.02$ and initial $S=E=F=0.3$. The last distribution remains stable
for much longer times.}
\end{figure}

In this standard network, we then use {\it directed} opinion links. This means
each person takes advice only from those four people whom the person
selected when joining the network; the same person ignores the opinions of
those who joined later and selected this person as advisor. Directed networks 
have been used before for, e.g., the Ising models \cite{sanchez} and opinion 
dynamics \cite{hmo}. A computer program was listed in \cite{sumour}.
Again, ordered sequential updating was used.

Figure 4 shows that in the present model everyone becomes normal: $S=E=F=0$ 
after sufficiently long time, differently from the results obtained with the
square lattice of the previous section. In contrast to the square lattices for 
$50 \times 50$ up to $20,000 \times 20,000$, larger networks needed
a slightly longer time for $E$ and $F$ to decay; see Fig. 5. With different 
parameters, also quite complicated dynamics can be found; see Fig. 6.

If we use symmetric instead of directed connections in these networks, now
containing only 100,000 people, then for $\gamma_1 = 0.1$ all four groups 
survive; for $\gamma_1=0.4$ the $F$ die out; for $\gamma_1 = 0.6$ also
the $E$ and for $\gamma_1 = 0.9$ we end up with $S=E=F=0$ (not shown).
Similar effects are seen if, initially, the first four people are fanatic and
all others have opinion $G$.

Thus, similarly to \cite{sumour} for Ising magnets and differently from 
\cite{hmo} for opinion dynamics, the directed network structure gives very 
different results compared to the undirected case. 

\section{Conclusions}

In summary, the model of \cite{bio} depends somewhat on the various 
changes in the underlying connections which we had introduced. 
Overall we feel that we confirmed the main conclusions from differential 
equations \cite{bio}: Depending on parameters like $\gamma_1$, fanatics
and/or excited agents survive or die out. 

DS thanks F. Bagheri-Tar for crucial help to survive in Los Angeles. 


\end{document}